\documentclass{aa}
\usepackage{graphicx}
\usepackage{epsfig}
\usepackage{graphics}
\begin{document}%
   \title{Gamma-ray emission from massive young stellar objects}

   \author{A.~T. Araudo\inst{1,2,}\thanks{Fellow of CONICET, Argentina},
           G.~E. Romero\inst{1,2,}\thanks{Member of CONICET, Argentina},
           V. Bosch-Ramon\inst{3}
           \and J. M. Paredes\inst{4}
          }

   \offprints{Anabella T. Araudo: \\ {\em aaraudo@fcaglp.unlp.edu.ar}}
   \titlerunning{Gamma-ray emission from massive YSOs}

\authorrunning{A.T. Araudo et al.}  \institute{Instituto Argentino de
Radioastronom\'{\i}a, C.C.5, (1894) Villa Elisa, Buenos Aires,
Argentina \and Facultad de Ciencias Astron\'omicas y Geof\'{\i}sicas,
Universidad Nacional de La Plata, Paseo del Bosque, 1900 La Plata,
Argentina \and Max Planck Institut f\"ur Kernphysik, Saupfercheckweg
1, Heidelberg 69117, Germany \and Departament d'Astronomia i
Meteorologia, Universitat de Barcelona, Mart\'{\i} i Franqu\`es 1,
08028, Barcelona, Spain}

\date{Received / Accepted}


  \abstract 
{Massive stars form in dense and massive molecular
  cores. The exact formation mechanism is unclear, but it is possible
  that some massive stars  are formed by processes similar to those
  that produce the low-mass stars, with accretion/ejection phenomena
  occurring at some point of the  evolution of the protostar. This
  picture seems to be supported by the detection of a collimated
  stellar wind emanating from the massive  protostar
  IRAS~16547$-$4247. A triple radio source is associated with the
  protostar: a compact core and two radio lobes. The emission of  the
  southern lobe is clearly non-thermal. Such emission is interpreted
  as synchrotron radiation produced by relativistic electrons locally
  accelerated at the termination point of a thermal jet. Since the
  ambient medium is determined by the properties of the molecular
  cloud in  which the whole system is embedded, we can expect high
  densities of particles and infrared photons. Because of the
  confirmed presence of  relativistic electrons, inverse Compton and
  relativistic Bremsstrahlung interactions are unavoidable.  }
{We aim at making quantitative predictions of the spectral energy
distribution of the non-thermal spots generated by massive young
stellar objects,  with emphasis on the particular case of
IRAS~16547$-$4247.}  
{We study the
high-energy  emission generated by the relativistic electrons that
produce the non-thermal radio source in IRAS~16547$-$4247.  We also
study the result of proton acceleration at the terminal shock of the
thermal jet and make estimates of the secondary gamma-rays and
electron-positron pairs produced by pion decay.}  
{We present spectral energy distributions for the southern
lobe of IRAS~16547$-$4247, for a variety of conditions. We show that
high-energy  emission might be detectable from this object in the
gamma-ray domain. The source may also be detectable at X-rays through
long exposures with current X-ray instruments.}  
{Gamma-ray telescopes like
GLAST, and even ground-based Cherenkov arrays of new generation can be
used to study non-thermal processes  occurring during the formation of
massive stars.}


\maketitle
%
\section{Introduction}

The mechanism of formation of massive stars remains one of the open
questions in the field of star formation. It is known that these stars
originate inside massive molecular clouds but the sequence of
processes that take place during the formation of the star are mostly
unknown. It has been suggested, for example, that the coalescence of
various protostars in the same cloud can lead to the emergence of a
massive star (e.g. Bonnell, Bate, \& Zinnecker 1998). Massive stars
appear in massive stellar associations where cloud fragmentation seems
to be common. Alternatively, a massive star could form by the collapse
of the core of a massive cloud, with associated episodes of mass
accretion and ejection, as observed in low-mass stars (e.g. Shu, Adams
\& Lizano 1987). In such a case, the effects of jets propagating
through the medium that surrounds the protostar should be detectable.

Recently, Garay et al. (2003) have detected a triple radio continuum
source associated with the protostar IRAS~16547$-$4247. The radio
source presents a linear structure consisting of a thermal core, and
two radio lobes. The southern lobe is clearly non-thermal, indicating
the presence of relativistic electrons that produce the observed
radiation by synchrotron mechanism. This non-thermal source has been
interpreted by Garay et al. (2003) as the termination point of one of
the jets ejected by the protostar. There, a strong shock would
accelerate the electrons up to relativistic energies by Fermi
mechanism (e.g. Bell 1978). The observed spectral index of
$\alpha\sim -0.6$ ($S_{\nu}\propto\nu^{\alpha}$) is in good agreement
with what is expected from an uncooled population of relativistic
electrons produced by diffusive shock acceleration at a strong
non-relativistic shock (e.g. Protheroe 1999).

The angular separation of the non-thermal source from the core
corresponds to a linear distance of only 0.14 pc, and then the
population of relativistic particles is inside the molecular
cloud. Hence, these particles are in a rich environment, with a high
density of ambient matter and photon field from the infrared
emission of the cloud. Inverse Compton (IC) and relativistic
Bremsstrahlung losses are then unavoidable for these particles. If
protons are accelerated at the termination shock along with the
electrons, then inelastic $pp$ collisions can take place, producing
pions, which will decay yielding gamma-rays, relativistic
electron-positron pairs and neutrinos. The radiation produced via all
these mechanisms will be likely steady at scales of years due to the
dynamical timescales of the processes occurring at the source.

The main goal of the present paper is to estimate the high-energy
yield of all these interactions, both leptonic and hadronic, in order
to ponder whether gamma-ray astronomy can be used to probe the massive
star formation and the outflows it could produce. Till now, thermal
radio and X-ray emission has been associated with the formation of
low-mass stars. Here we will show that massive protostars can produce
a significant amount of radiation in the gamma-ray domain, because of 
the dense and rich medium in which they are formed.

The structure of the paper is as follows. In the next section we
provide the basic information about IRAS~16547$-$4247, the associated
radio sources, and the ambient medium. Then, in Section~\ref{loss} we
discuss the particle acceleration and the different losses for the
relativistic particles in the southern lobe of the radio
source. Section~\ref{gamma} deals with the gamma-ray production. Our
results are there presented in the form of spectral energy
distributions (SEDs), for different sets of parameters. We close with
a brief discussion and a summary in Section~\ref{disc}.

\section{The protostar IRAS~16547$-$4247 and its host environment}

The source IRAS~16547$-$4247 corresponds to a young massive
star-forming region, associated with an O-type protostar, located at
2.9~kpc (Garay et al. 2003). The luminosity of the source is $L \sim
6.2\times 10^{4} L_{\odot}\approx 2.4 \times 10^{38}$~erg~s$^{-1}$,
peaking at the infrared, which makes of it the most luminous detected
young stellar object (YSO) with thermal jets. Brooks et al. (2005)
have reported the detection of a chain of H$_{2}$~2.12~$\mu$m emission
knots that trace the collimated outflow that emanates from the center
of the source. The SED of the IR emission can be described by a
modified blackbody function with a peak temperature of 30~K. The total
mass of the cloud is $M_{\rm cl}=9\times 10^{2}$~$M_{\odot}$ (Garay et
al. 2003). Molecular line observations indicate that the size of the
cloud is $\sim0.38$~pc in diameter ($\approx 1.1\times
10^{18}$~cm). If we assume a spherical geometry, the averaged particle
(atoms of H) density of the cloud is $n_{\rm cl}\approx5.2 \times
10^{5}$~cm$^{-3}$. The energy density of IR photons in the cloud,
assuming an homogeneous distribution, is $w_{\rm ph}\approx 1.8 \times
10^{-9}$~erg~cm$^{-3}$.

\begin{figure}
\includegraphics[angle=0, width=0.4\textwidth]{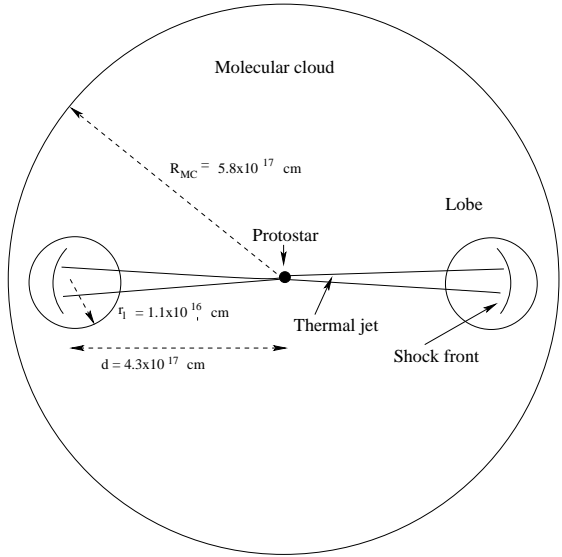}
\caption{Sketch of the scenario discussed in this work.}\label{fig_1}
\end{figure}

The radio observations made by Garay et al. (2003) with the ATCA and
the deeper observations by Rodr{\'\i}guez et al. (2005) with the VLA
show the existence of a triple radio source inside the molecular
cloud. The three components of the radio source are aligned in the
northwest-southeast direction, with the outer lobes separated from the
core by a projected distance of 0.14~pc. The central source is
elongated and has a spectral index of $0.33\pm0.05$, consistent with
free-free emission from a collimated jet (Rodr{\'\i}guez et
al. 2005). The radio lobes have some substructure. The integrated
emission from the northern lobe has a spectral index of $-0.32\pm
0.29$, of dubius thermal/non-thermal nature. The spectrum of the 
southern lobe radiation, instead, has an index
$\alpha=-0.59\pm 0.15$. Actually, the shock heated material can radiate
enough as to ionize the surrounding medium and free-free absorption could
modify to some extent the radio spectrum. However, we have no enough data to
characterize the region in order to derive its thermal properties or the
ionization degree. Otherwise, diffusive shock acceleration by strong
non-relativistic shocks naturally produces a power-law particle distribution
with an index similar to what is inferred from the observations. Thus, we
take as a first order approximation the observed radio spectrum as the
original one, which would correspond to a non-thermal particle
distribution.
 The inferred linear size for this lobe is $\approx 1.1\times 10^{16}$~cm.
 Non-thermal radio emission has been associated in the past with
the outflows of a few massive YSO (see Rodr{\'\i}guez et al. 2005, and
references therein). In any case the molecular cloud is so massive and
luminous as in the case of IRAS~16547$-$4247. The flux density of the
southern lobe is $2.8\pm 0.1$~mJy at 8.46~GHz.
In Figure~\ref{fig_1} a sketch of the scenario discussed in this paper 
is shown.

\section{Particle acceleration and losses}\label{loss}

At the termination point of the thermal jet a strong shock front is
expected to be formed. Charged particles can be accelerated at the
shock,  with an acceleration rate $\propto \eta B$ (see below), where
$B$ is the magnetic field and the efficiency $\eta$ depends on the
acceleration mechanism and its details. In the case of diffusive shock
acceleration $\eta\sim (v_{\rm s}/c)^2/f_{\rm sc}$, where $v_{\rm s}$
is the shock velocity and $f_{\rm sc}$ is the ratio of the mean free
path of particles to their gyro-radius. Close to the Bohm limit
$f_{\rm sc}\sim 1$. The terminal velocity of the collimated outflow in
IRAS~16547$-$4247 is unknown, but Mart{\'\i}, Rodr{\'\i}guez \&
Reipurth (1995) have determined a velocity in the range
$600-1400$~km~s$^{-1}$ for the HH 80-81 thermal radio jet, which is
powered also by a massive YSO. Adopting a value of $\sim 1000$ km
s$^{-1}$ for IRAS~16547$-$4247, we have a reasonable efficiency of
$\eta\sim10^{-5}$ for the southern lobe of this source. The efficiency
in the northern lobe would be lower, since the non-thermal emission
does not seem to be dominant there.

The average matter density of the cloud is $5.2 \times
10^{5}$~cm$^{-3}$. It is noted nevertheless that in the shocked
material the density should be $\sim 4$ times this value. In our
calculations of the Bremsstrahlung and pion-decay radiation we have 
adopted this higher density.

As mentioned above, the rate of energy gain for electrons at the 
acceleration region is:
\begin{equation}
\dot{\gamma}_{e, \;\rm{gain}} = \frac{\eta eBc}{m_e c^2},
\end{equation}  
where $\gamma$ is the Lorentz factor of the particle, and $m_e$ is the
electron rest mass.  Similarly, for protons,
\begin{equation}
\dot{\gamma}_{p, \;\rm{gain}} = \frac{\eta eBc}{m_p
c^2}. \label{gain-p}
\end{equation} 

\subsection{Non-thermal radiation losses and maximum energies}

In order to get the maximum energy of the primary particles we have to
balance the energy gain and loss rates. In the case of electrons,  the
relevant losses are synchrotron, IC, and relativistic Bremsstrahlung
losses. The latter two, under the conditions described in the previous
section,  are given by:
\begin{equation}
\dot{\gamma}_{\rm{IC}} = -3.2\times10^{-17}\gamma^2\;\;\rm{s^{-1}},
\end{equation}
\begin{equation}
\dot{\gamma}_{\rm{Br}} = -7.4\times10^{-10}\gamma\;\;\rm{s^{-1}}.
\end{equation}   
The expression for the synchrotron losses depends on the magnetic
field $B$ and is given below. We can estimate the magnetic field
assuming equipartition in energy density between field and particles 
in the radio lobe, so that:
\begin{equation}  
\frac{B^2}{8\pi} = u_{\rm{e_1}} + u_p + u_{\rm{e_2}}.
\end{equation}
Here, $u_{\rm{e_1}}$ is the energy density of primary electrons,
$u_{\rm p}$ that of primary protons, and $u_{\rm{e_2}}$ corresponds to
secondary pairs that will result from the charged pions produced in
inelastic $pp$ collisions. In each case  ($i = {\rm{e_1}}, p,
{\rm{e_2}}$):
\begin{equation}  
u_{i}=\int E_i n(E_i) dE_i,
\end{equation}
where $n(E_i)=K_i E_i^{-\Gamma_i}\exp(-E_i/E_i^{\rm{max}})$ is the
particle density distribution [cm$^{-3}$~erg$^{-1}$].

It is a fact that there are relativistic electrons or pairs in the
southern lobe. This population of relativistic leptons can be produced
via acceleration either of electrons or protons. In the former case,
the accelerated electrons would be the responsible for the production
of the detected synchrotron emission. In the latter case, accelerated
protons would suffer $pp$ interactions with nuclei of the ambient
medium generating charged pions that would decay to muons, and these
subsequently decaying to pairs. Then, these pairs would be the
emitters of the observed non-thermal radiation. Although the calculation of 
the neutrino emission is out of the scope of this work, we note that
neutrino luminosities would be similar to those of $\pi^0$-decay gamma-rays.

The most likely scenario is one with acceleration of both electrons and
protons. Nevertheless, the relative number of accelerated protons is
unknown, so we will write $u_p = au_{\rm{e_1}}$ and then
$u_{\rm{e_2}}=f u_p$. 
The value of $f$ was estimated using the average ratio of the number 
of secondary pairs to $\pi^0$-decay 
photons (Kelner, Aharonian \& Bugayov, 2006). 
We will consider three cases: $a=0$
(no proton acceleration), $a=1$ (equal energy density in protons as in
electrons), and  $a=100$ (proton dominance, as it is the case in the
galactic cosmic rays). 
 
For a distribution of protons $n(E_p)$, the energy spectra of secondary 
electron-positron pairs was calculated using the new parametrization of 
the inelastic cross-section of $pp$ interactions given by Kelner et al. 
(2006):  
\begin{equation}\label{cross-section}
\sigma_{\rm{inel}}(E_p) = 34.3 + 1.88L + 0.25L^2 \;\;\rm{mb},
\end{equation}                                                                 %
where $L = \ln(E_p/1\;\rm{erg})$.                                                                                                                             
In the cases $a=0$ and $a=1$, $\Gamma_{\rm{e_1}}=2.18$, from the radio
observations. For $a=1$, when primary electrons are still dominant,
the same slope is assumed for them and for protons. Then, for the
secondaries, we get $\Gamma_{\rm{e_2}}=2.13$.  In the case of $a=100$,
where the secondary leptonic emission dominates,
$\Gamma_{\rm{e_2}}$ must be $2.18$ and then $\Gamma_p=\Gamma_{\rm{e_1}}=2.27$
for the primary particles.

Using the standard synchrotron formulae (Ginzburg \& Syrovatskii, 1964) 
and the data from
Rodr{\'\i}guez et al. (2005), along with the equipartition condition
and the equations given above, we get the magnetic fields and the
normalization constants shown in Table~\ref{k_B}, for each case
considered. Typically, $B$ is about few $10^{-3}$~G. 

\begin{table*}[]
\begin{center}
\caption{Magnetic field and normalization constants for the different cases considered in the text.}\label{k_B}
\begin{tabular}{ccccc}
\hline
a  & B   & $K_{\rm{e_1}}$ & $K_{\rm p}$ & $K_{\rm{e_2}}$ \\
{} & [G] & $[\rm{erg^{\Gamma_{\rm{e_1}}-1}cm^{-3}}]$ & $[\rm{erg^{\Gamma_{\rm p}-1}cm^{-3}}]$ & $[\rm{erg^{\Gamma_{\rm{e_2}}-1}cm^{-3}}]$\\
\hline
0 & $2.0\times10^{-3}$ & $2.6\times10^{-9}$ & - & - \\
1   & $2.5\times10^{-3}$ & $2.0\times10^{-9}$ & $6.7\times10^{-9}$ & $6.5\times10^{-10}$ \\ 
100  & $3.0\times10^{-3}$ & $2.7\times10^{-11}$ & $1.7\times10^{-8}$ & $1.5\times10^{-9}$  \\ 
\hline
\end{tabular}
\end{center}
\end{table*}  
    
Now, with this magnetic field we can calculate the synchrotron losses in each case:
\begin{equation}
\dot{\gamma}_{\rm{synchr}}(a=0) = -1.0\times10^{-14}\;\gamma^2\;\;\rm{s^{-1}},
\end{equation}
\begin{equation}
\dot{\gamma}_{\rm{synchr}}(a=1) = -1.1\times10^{-14}\;\gamma^2\;\;\rm{s^{-1}},
\end{equation}
\begin{equation}
\dot{\gamma}_{\rm{synchr}}(a=100) = -3.3\times10^{-14}\;\gamma^2\;\;\rm{s^{-1}}.
\end{equation}     

\begin{figure}
\includegraphics[angle=270, width=0.5\textwidth]{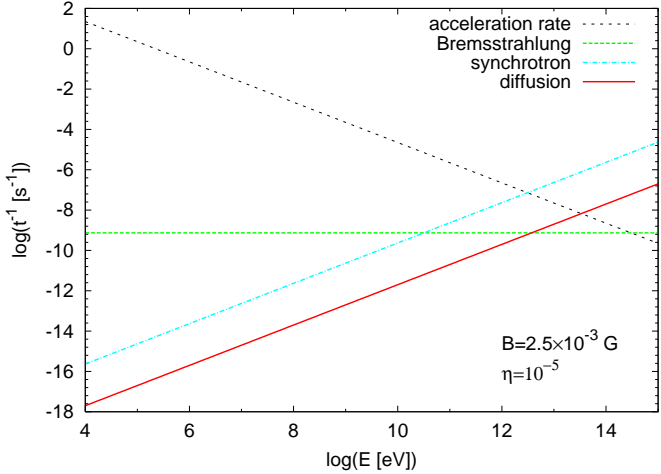}
\caption{Energy loss rates for synchrotron and relativistic Bremsstrahlung 
processes. The IC energy loss rate is not shown since it is negligible.
The acceleration rate and the particle rate of diffusion 
out of the emitting region are also shown.}\label{losses}
\end{figure}

The different losses for electrons are shown in Fig.~\ref{losses},
along with the acceleration rate, for a value of $B$ of
$2.5\times10^{-3}$ G (case $a=1)$. We can see that low energy
electrons cool mainly through relativistic Bremsstrahlung, whereas the
losses of the most energetic ones are dominated by synchrotron
radiation. The dominant cooling regime changes at $E_e=E_{\rm b}$,
where $E_e$ is the electron energy, either of primaries or
secondaries. For a steady state, under the conditions  considered
here, the electrons will have an index
$\Gamma_{{\rm{e_1}},{\rm{e_2}}}$ for energies $E_e< E_{\rm b}$ and
$\Gamma_{{\rm{e_1}},{\rm{e_2}}}+1$ for $E_e> E_{\rm b}$. In Table
\ref{break} we show the values of $E_{\rm b}$ for the different cases
under consideration, along with the corresponding normalization
constants at high energies.

\begin{table}[]
\begin{center}
\caption{Break energy in the electron spectra for different cases
discussed in the text. Normalization constants  for the power-law
distributions above the break energy are shown as well.}\label{break}
\begin{tabular}{cccc}
\hline 
a    & $E_{\rm b}$ & $K_{\rm{e_1}}'$ & $K_{\rm{e_2}}'$ \\ {} &
[eV] & $[\rm{erg^{\Gamma_{\rm{e_1}}}cm^{-3}}]$ &
$[\rm{erg^{\Gamma_{\rm{e_2}}}cm^{-3}}]$\\ 
\hline 
$0$   & $9.4\times10^{9}$  & $3.7\times10^{-11}$ & - \\  
$1$   & $3.1\times10^{10}$ & $9.8\times10^{-11}$ & $3.2\times10^{-11}$\\
$100$ & $2.2\times10^{10}$ & $9.5\times10^{-13}$ & $5.4\times10^{-11}$ \\
\hline
\end{tabular}
\end{center}
\end{table}

The relevant losses for protons are from inelastic $pp$ collisions
with the cloud material. In our case, these losses are approximately given 
by (Mannheim \& Schlickeiser 1994):
\begin{equation}
\dot{\gamma}_{pp} = -9.2\times10^{-10}\left[0.95 +
0.06\ln\left(\frac{\gamma}{1.1}\right)\right]\gamma\;\;\rm{s^{-1}},
\end{equation}
where now $\gamma$ is the Lorentz factor of the protons.  Equating
with Eq.~(\ref{gain-p}), we get the maximum energy of the accelerated
protons, which results of the order of  $\sim 10^{15}$~eV. Although
the gyro-radius of these particles can be contained in the
acceleration region, they can diffuse  out of the cloud before
reaching the mentioned maximum energy, hence a more tight constraint on the
maximum energy can be obtained  from the following equation:
\begin{equation}
\frac{E_p}{\eta eBc} = \frac{r_{\rm lobe}^2}{D},
\end{equation}
where $D$ is the diffusion coefficient and $r_{\rm lobe}\approx
1.1\times10^{16}$ cm the size of the acceleration region. This assumption
 is supported by the fact that particles lose all their energy before
they diffuse out of the emitting region 
(i.e. $\tau_{\rm cooling} < \tau_{\rm diff}$) (see Fig.~\ref{losses}).
Working on
the Bohm limit we can set $D=D_{\rm B}=r_{\rm g} c/3$, with $r_{\rm
g}$ the gyro-radius.  In such a case we get the maximum energies shown
in Table~\ref{diffusion}, which are more than one order of magnitude
lower than in the previous calculation.

\begin{table}[]
\begin{center}
\caption{Maximum energies obtained for electrons and protons
accelerated in the southern lobe of the radio source associated with
IRAS~16547$-$4247. The  maximum energy for secondary pairs is shown in
the last column.}\label{diffusion}
\begin{tabular}{cccc}
\hline
a  & $E_{\rm{e_1}}^{\rm{max}}$ & $E_{\rm p}^{\rm{max}}$ & $E_{\rm{e_2}}^{\rm{max}}$ \\
{} & [eV]                 & [eV]             & [eV]                 \\
\hline
$0$   & $3.1\times10^{12}$ & -                  & -                   \\
$1$   & $3.1\times10^{12}$ & $4.7\times10^{13}$ & $1.1\times10^{12}$ \\ 
$100$ & $2.8\times10^{12}$ & $5.7\times10^{13}$ & $1.8\times10^{12}$ \\
\hline
\end{tabular}
\end{center}
\end{table}

\subsection{Relativistic particle distribution}

 The lepton particle distribution in the emitting region can be calculated, 
as a
first order approximation, adopting a one-zone model, in which particles, once
injected following a power-law energy distribution, evolve suffering mainly
synchrotron and relativistic Bremsstrahlung losses. In such a context, the
spectrum of the particles is determined by the following transport equation
(e.g. Khangulyan et~al. 2007):
\begin{equation}
\partial n\left(t,\gamma\right)/\partial t+\partial \dot{\gamma} 
n\left(t,\gamma\right)/\partial\gamma+n\left(t,\gamma\right)/
\tau_{\rm esc}=Q(t,\gamma),
\label{Ginz_equation}
\end{equation}
where $Q(t,\gamma)$ is the function for the particle injection, which
takes place during the lifetime of the source ($\tau_{\rm life}$), 
estimated in $\ga
100$~yr$\approx 3\times 10^9$~s (see Garay et~al. 2003). 
Here, $t$ is the
time, and $\gamma$ is the lepton Lorentz factor. 
Admittedly, the age of the source is not well constrained, although 
it could hardly be significantly younger than $\sim 100$~yr, the minimum 
jet crossing  time, since the shock has been probably active for most of 
this period while the shocked material was being \emph{displaced} to its
 present location.
Concerning $\tau_{\rm
esc}$, at this stage we take it as being the shortest diffusion
timescale, i.e. that corresponding to the maximum energy particles. 
The time-derivative $\dot{\gamma}$ is a function 
accounting for all the energy losses affecting leptons, 
i.e. basically synchrotron and relativistic Bremsstrahlung.
The solution of Eq.~(\ref{solution0}) is:
\begin{equation}
n(t,\gamma)=\frac{1}{\dot{\gamma}}\int\limits_{\gamma}^{\gamma_{\rm eff}} 
Q(t-\tau,\gamma'){\rm e}^{-\tau({\gamma,\gamma'})/\tau_{\rm esc}}
d\gamma',
\label{solution0}
\end{equation}
where
\begin{equation}
t=\int\limits_{\gamma}^{\gamma_{\rm eff}}\frac{d\gamma'}{\dot{\gamma'}}\;\;\;{\rm
and}\;\;\;
\tau(\gamma,\gamma')=\int\limits_{\gamma}^{\gamma'}
\frac{d\gamma''}{\dot{\gamma''}}.
\label{tt}
\end{equation}
The time dependence comes through Eq.~(\ref{tt}). For small $t$, 
$\gamma\la \gamma_{\rm eff}$; for larger $t$, $\gamma\ll\gamma_{\rm eff}$. 
Since $\gamma_{\rm eff}$ 
can be arbitrarily large, for $\gamma'$ in 
Eq.~(\ref{solution0}) above the maximum Lorentz factor
$Q$ will be zero. In addition, if $\tau$, 
being shorter than $\tau_{\rm cooling}$, becomes 
larger than $\tau_{\rm esc}$, the final
particle spectrum will be affected by particle escape.

\begin{figure*}
\centering
\includegraphics[angle=270, width=0.4\textwidth]{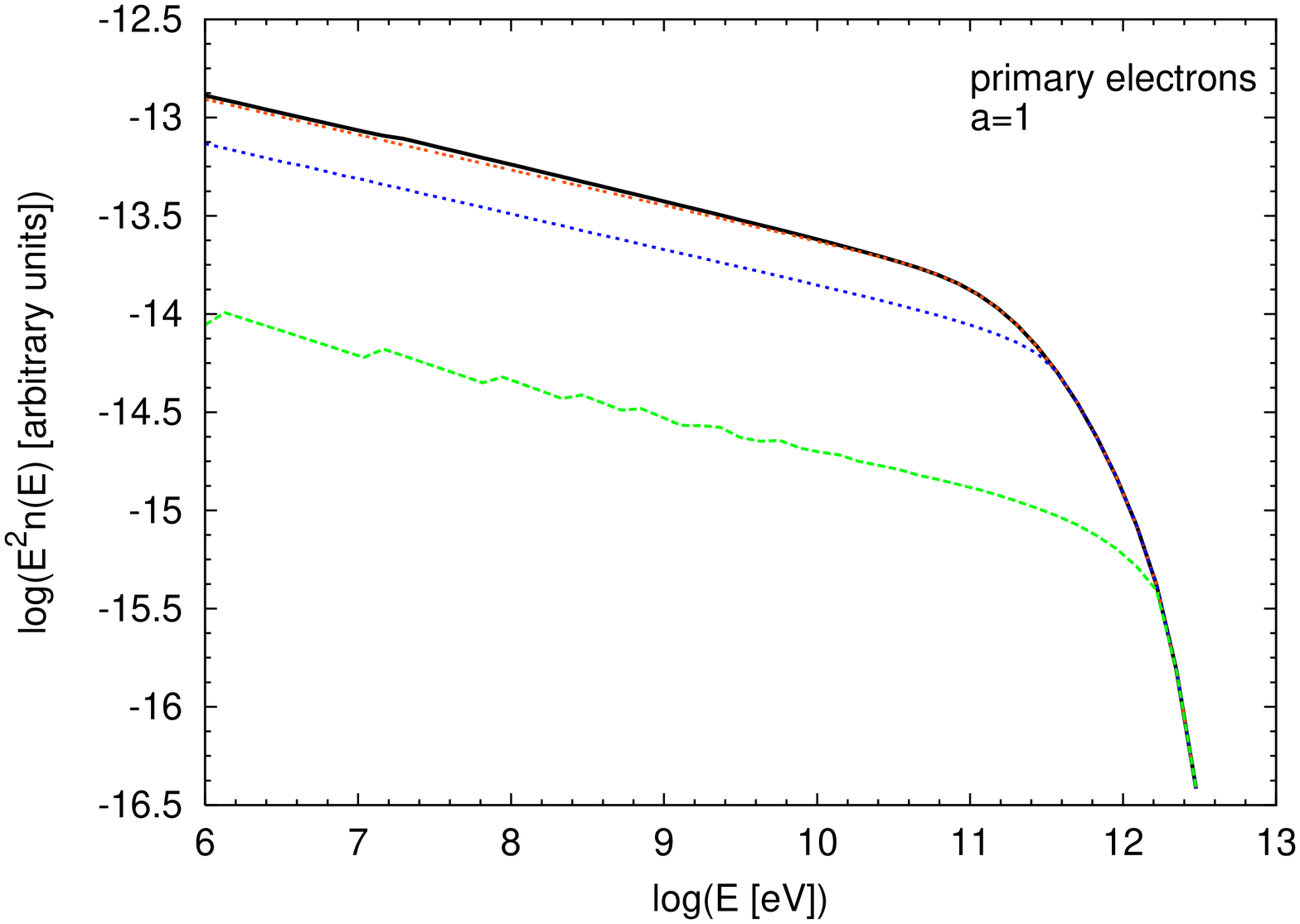}\qquad
\includegraphics[angle=270, width=0.4\textwidth]{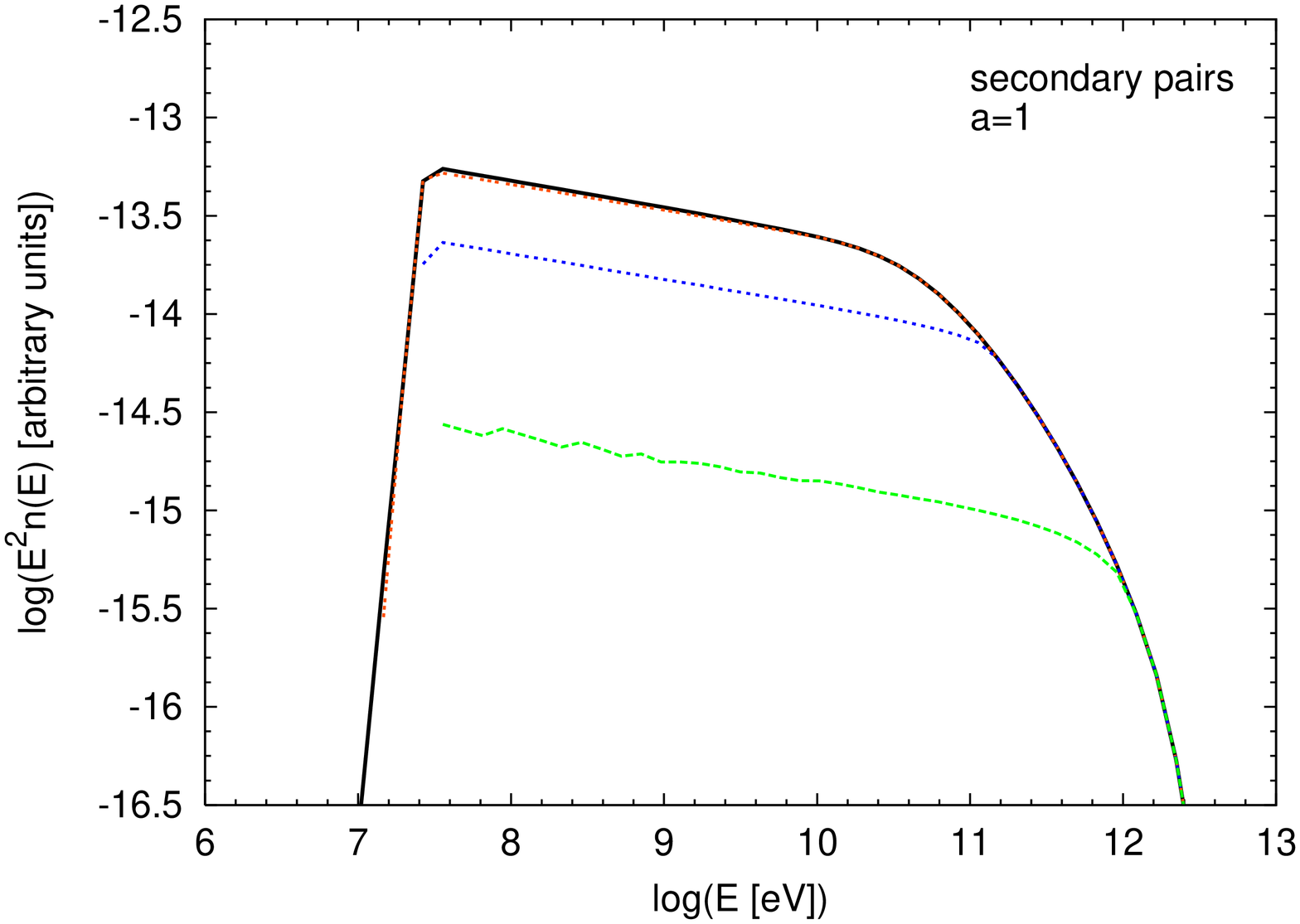}\\[10pt]
\includegraphics[angle=270, width=0.4\textwidth]{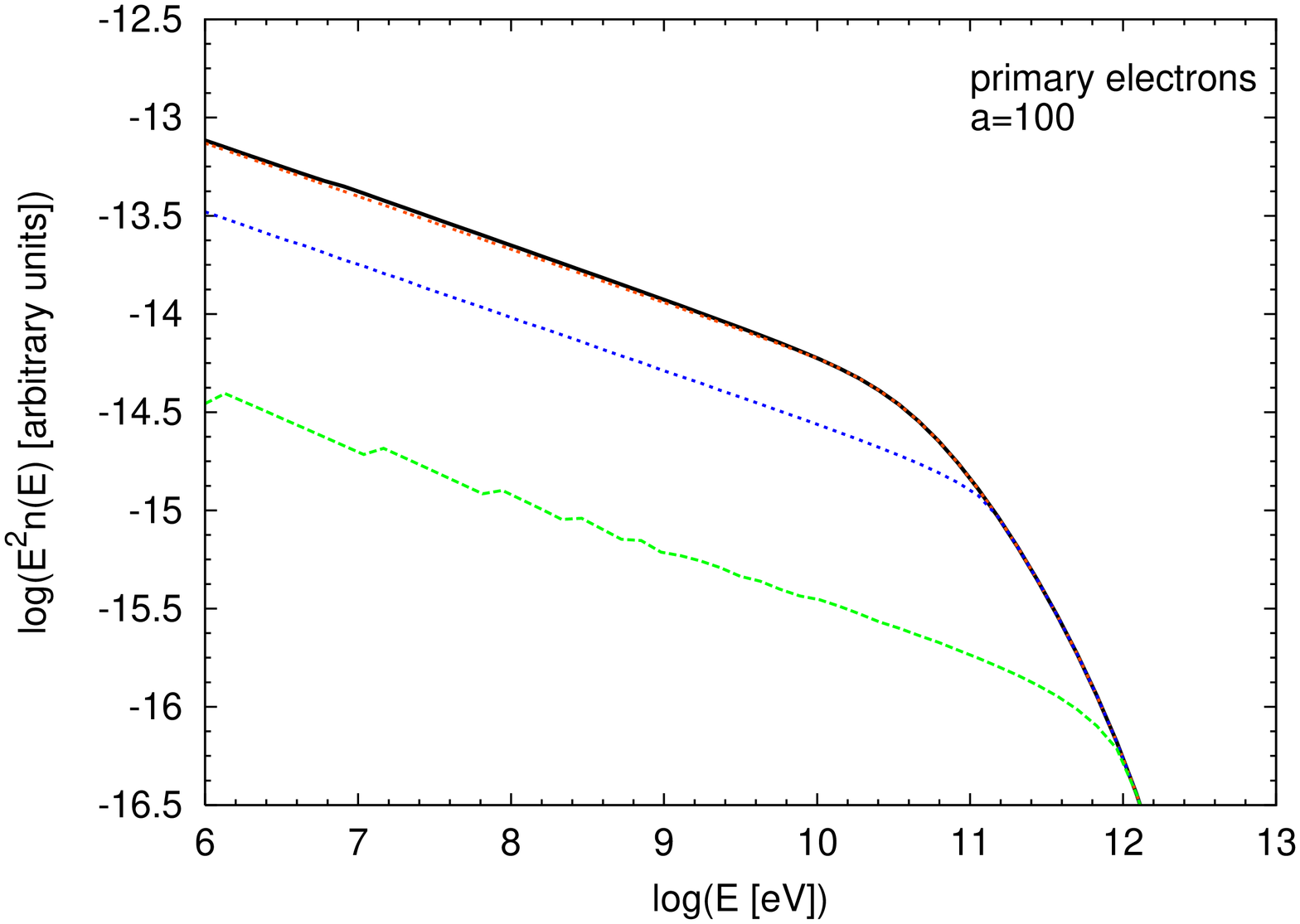}\qquad
\includegraphics[angle=270, width=0.4\textwidth]{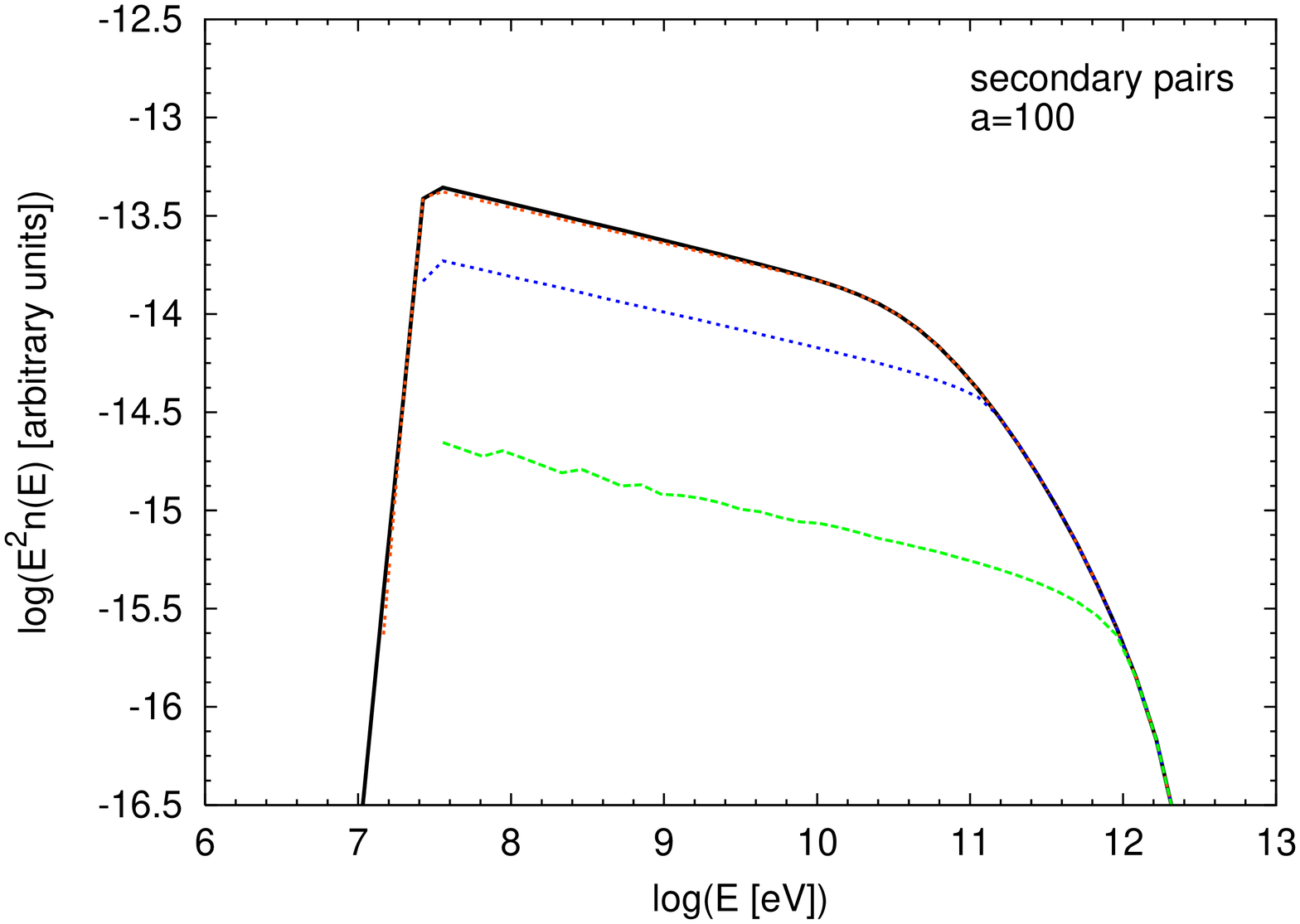}\\
\caption{Spectral energy distributions of primary electrons and secondary 
pairs.
Distributions for different particle injection timescales ($\tau_{\rm inj}$) 
are shown in each panel. Different curves correspond to the following values of
$\tau_{\rm inj}$: $10^7$ s (long-dashed line, green), $10^8$ s (dot-dashed 
line, blue), $10^9$ s
(dotted line, red) and $10^{10}$~s (solid line, black). 
Notice that for ages $> 10^9$~s the
steady regime is reached (the curves appear in color only in the on-line 
version of the paper).}
\label{evol}
\end{figure*}

The computed particle energy distributions of primary and secondary leptons  are
shown in Fig.~\ref{evol}. As seen in the figure, for injection  timescales
$\tau_{\rm{inj}}\sim 10^9$~s$< \tau_{\rm life}$,  the particle distribution $n\left(t,\gamma\right)$ has
already reached the steady regime, as expected from the fact that $\tau_{\rm
cooling}\sim 10^9$~s.  Moreover, the final particle energy distribution is not
affected by $\tau_{\rm esc}$ (which is actually a lower limit for the timescale
of diffusion particle
escape), since $\tau_{\rm esc}\ga \tau_{\rm cooling}$ (see Fig.~\ref{losses}), 
i.e. particles will radiate inside the emitting region.

Regarding protons, their diffusion timescales and source lifetime are
long enough to let them radiate almost all their energy in the hot
spot. Moreover, the energy loss timescale is only slightly dependent
on energy. This implies that proton energy distribution, in the
context of our scenario, keeps almost the same spectral shape as that
of the injected one, and reaches, as it is the case for leptons, the
steady  regime, without suffering significant impact from escape
losses.

\subsection{Thermal radiation and ionization losses in the shocked region}

 In the context of an outflow interacting with its environment, a
strong shock of several hundreds~km~s$^{-1}$ can heat the shocked
cloud material up to temperatures of $\sim 10^7$~K, producing emission
that would peak at soft X-rays. The heating of the medium would take
place via shock compression, and ionization and Coulomb scattering of
the thermal atoms/ions would occur due to the presence of relativistic
leptons and protons. These soft X-rays should be strongly absorbed by
the surrounding medium, provided the large hydrogen column density of
the cloud. Hence, this radiation will be hard to detect. In any case,
this would lead to ionization of the cold material surrounding and/or
embedded by the emitting region, which may affect to some extent the
radio emission via free-free absorption. Since the observed radio
spectrum seems to be non-thermal and we aim at obtaining just first
order estimates of the high-energy emission, we assume here that the
overall effect of the free-free absorption is not dominant in the
southern lobe of the source. Nevertheless, a more detailed
characterization of the original radio spectrum would deserve further
study.  Another observational consequence of the ionization of the
medium would be the detection of recombination lines, although their
study is beyond of the scope of our work.

It is worth mentioning that, if the emitting region were not
significantly ionized by the jet termination shock, ionization losses
would affect the particle energy distribution. The effect of this loss
channel, which energy loss rate does not depend on energy, would be to
harden the electron energy distribution at the energies where
ionization losses are greater than those due to relativistic
Bremsstrahlung, i.e. at Lorentz factors $\gamma< 10^3$. We note that, 
if energy losses were indeed dominant at $\gamma< 10^3$,
their impact would be a pretty hard radio spectrum, which is apparently 
not the case.

\section{Production of gamma-rays and lower energy radiation}\label{gamma}

From the previous section it is clear that at the termination shock
there will be particles energetic enough as to produce gamma-ray
emission. In this section we will use the particle distributions, 
ambient photon density and magnetic fields already estimated 
to calculate the non-thermal spectral energy distributions expected
from the different particle interactions in the southern lobe of the radio
source associated with IRAS~16547$-$4247.

\subsection{Leptonic interactions}

\begin{figure}
\includegraphics[angle=270, width=0.5\textwidth]{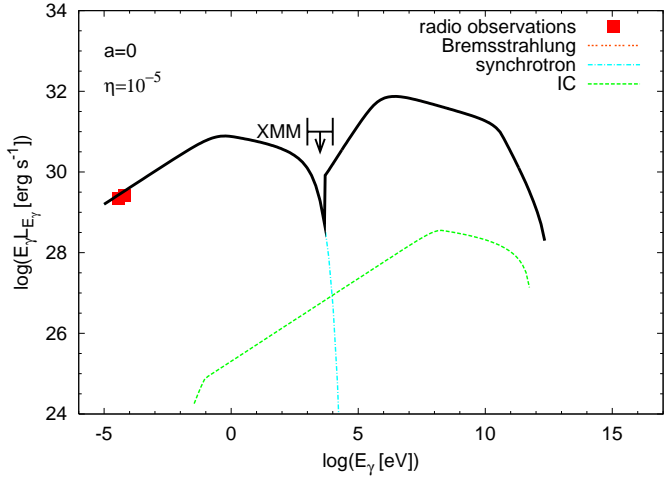}
\caption{Spectral energy distribution for a pure leptonic case.}\label{SED1}
\end{figure}

Synchrotron, relativistic Bremsstrahlung and IC emissivities for
primary and secondary leptons were calculated using the standard
formulae (e.g.  Blumenthal \& Gould 1970, Pacholczyk 1970). Full
Klein-Nishina cross section was used for the IC computation. The
emission region was considered spherical, with a radius $r = r_{\rm
lobe}$.  Figure~\ref{SED1} shows the results for the pure leptonic
case ($a=0$).

Focusing on the energy bands at which non-thermal emission may be
detectable, we see that the high-energy emission is dominated by the
relativistic Bremsstrahlung, with a peak of $\sim
10^{32}$~erg~s$^{-1}$ at $E_{\gamma}\sim 1$~MeV. At energies
$E_{\gamma}\ga 1$ GeV, the source presents luminosities of $\sim 10^{31}$
erg s$^{-1}$, with a cutoff $\sim 10$~GeV. At X-rays, in the range
1--10~keV, the expected luminosities are about $10^{30}$~erg~s$^{-1}$,
presenting a softening of the spectrum due to the high energy cutoff of 
the electron energy distribution.

\subsection{Hadronic interactions}

In the cases with $a=1$ and $a=100$, protons will have a significant
contribution. The differential $\gamma$-ray emissivity generated
through $\pi^0$-decay was calculated as  

\begin{equation}
q_{\gamma}(E_{\gamma})=2\int^{\infty}_{E_{\pi}^{\rm min}(E_\gamma)}
\frac{q_{\pi^0}(E_{\pi})}{\sqrt{E_{\pi}^{2}- m_{\pi}^{2} c^4}}
\;dE_{\pi}, \label{qgama}
\end{equation}  
where $E_{\pi}^{\rm min}(E_{\gamma})=E_{\gamma}+\frac{m_{\pi}^{2}
c^4}{4E_{\gamma}}$. Applying the $\delta$-function approximation for
the differential cross section\footnote{This approximation considers
only the most energetic neutral pion that is produced in the $pp$
reaction aside of a {\em fireball} composed by a certain number
of less energetic $\pi$-mesons of each flavor. See
a discussion in Pfrommer and En$\beta$lin (2004).}(Aharonian \&
Atoyan, 2000), the pion emissivity becomes
\begin{eqnarray}\label{q_pi}
q_{\pi^0}(E_{\pi})&= 4 \pi \int_{E_{\rm th}}^{\infty}
\delta(E_\pi-\kappa{E_{\rm kin}}) J_p(E_p)\,\sigma_{pp}(E_p)
\,dE_p\nonumber \\ &=\frac{4\pi}{\kappa} J_p\left( m_p
c^2+\frac{E_{\pi}}{\kappa}\right)\sigma_{pp}\left( m_p
c^2+\frac{E_{\pi}}{\kappa}\right)
\end{eqnarray}
for proton energies greater than the energy threshold $E_{\rm
th}=1.22$ GeV and lower than $100$ GeV. Here, $\kappa$ is the mean fraction of the kinetic
energy $E_{\rm kin}=E_p-m_p c^2$ of the proton transferred to a
secondary meson per collision. For a broad energy region (GeV to TeV)
$\kappa\sim 0.17$. The total cross section of the inelastic $pp$
collisions is  given by Eq.~(\ref{cross-section}). In Eq.~(\ref{q_pi}) $J_p(E_p) = (4\pi/c)n(E_p)$ is the proton flux. For $E_p > 100$ GeV, the equations given by Kelner et al (2006) were used.

The specific luminosity can then be estimated as:
\begin{equation}
E_{\gamma}L_{E_{\gamma}}=E_{\gamma}^{2}\int 	q_{\gamma}(E_{\gamma}) n({\bf{r}}) d{\bf{r}},
\end{equation}
where $n({\bf{r}})$ is the ambient particle density.

\begin{figure}
\includegraphics[angle=270, width=0.5\textwidth]{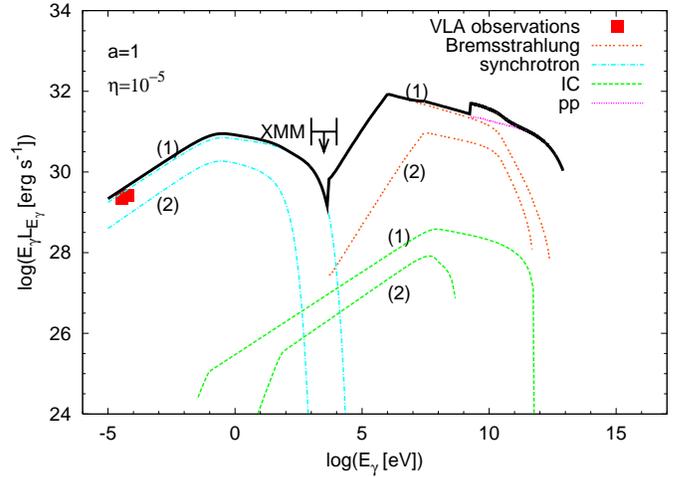}
\caption{Spectral energy distribution for a mixed primary population of 
relativistic electrons and protons. We indicate with numbers (1) and (2) 
the contribution from primary and secondary electrons, respectively.}\label{SED2}
\end{figure}

\begin{figure}
\includegraphics[angle=270, width=0.5\textwidth]{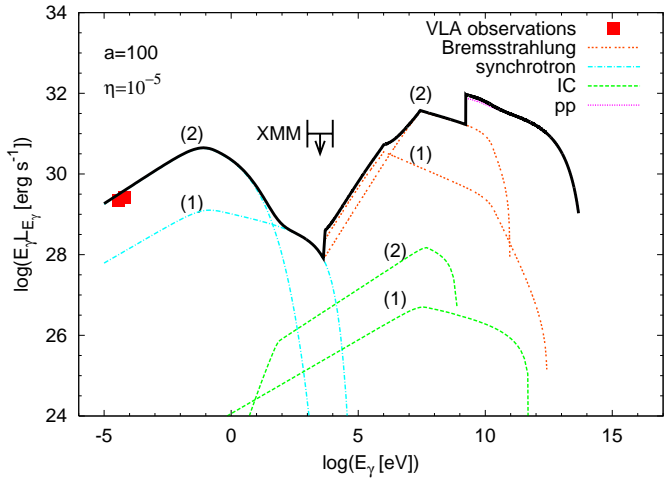}
\caption{Idem Fig.~\ref{SED2}, but for a case dominated by relativistic protons.}\label{SED3}
\end{figure}

Results from our calculations can be seen in Figures~\ref{SED2} and
\ref{SED3}. The hadronic emission is similar in luminosity to the
relativistic Bremsstrahlung ($\sim 10^{32}$ erg s$^{-1}$), but extends
up to higher energies $\sim 1$ TeV. Primary electrons dominate the
synchrotron and relativistic Bremsstrahlung emission in the case
$a=1$, but for $a=100$ (see Fig. 4) the secondary leptons produce the most
important contribution. In neither case the IC plays a significant
role. Basically, the emission above $E_{\gamma}\sim 1$ GeV is due to
neutral pion decays, whereas the X-ray and soft gamma-rays are
produced by relativistic Bremsstrahlung. For the cases $a=1$ and 100,
the synchrotron and relativistic Bremsstrahlung components produced by
secondary pairs are similar in spectral shape to those generated
by primary electrons in the same cases, but with different  behavior
at low energies, due to the low-energy cutoff introduced by the
minimum energy at which secondaries are injected.

\section{Discussion and summary}\label{disc}

Gamma-ray sources on the galactic plane are usually associated with
star forming regions (Romero, Benaglia \& Torres 1999). However, the
sources detected by EGRET are far more luminous than the sources that
can be produced by massive YSOs according to our calculations. EGRET
unidentified sources are most likely produced by pulsars, supernova
remnants, massive stars (Romero 2001), and microquasars (Bosch-Ramon,
Romero \& Paredes 2005). The emission of massive YSOs falls below the
sensitivity of EGRET, but they can be detected in principle by
GLAST. At very high energies ($E_{\gamma}\ga 100$ GeV), Cherenkov
arrays like HESS-II could be able to detect a source like
IRAS~16547$-$4247. Since such instruments might find the high-energy
cutoff, valuable information could be obtained about the true
efficiency of the acceleration process.

At X-rays, the source may be detectable by {\it Chandra} and {\it
XMM-Newton}, probably as point-like, through a deep observation of the
region.  Briefly looking at archive {\it XMM-Newton} 30~ks
observations of IRAS~16547$-$4247, there is not significant emission
above the background from the southern lobe region. This yields an
upper-limit for the 1--10~keV luminosity of $\sim
10^{31}$~erg~s$^{-1}$ at the distance of 2.9~kpc. This value is not in
contradiction with the luminosities of synchrotron and the
relativistic Bremsstrahlung components expected in this energy
range (see Figs.~\ref{SED1},  \ref{SED2} and
\ref{SED3}). Nevertheless, a deeper observation may detect radiation
from this region. In case the temperatures of the shocked material
 were high enough, a thermal component may be
detectable in the X-rays as well. In the UV-band, the
luminosity may still be higher than that produced by non-thermal
particles, but could be difficult to detect.

An issue that is relevant for the feasibility of the scenario
discussed in this paper is that the non-thermal luminosities are well
below the kinetic energy budget of the jet in IRAS~16547$-$4247. The
jet mass-loss rate has been estimated by Garay et al. (2003) in $\sim
10^{-5}$~M$_{\odot}$~yr$^{-1}$. With a velocity of $\sim
10^3$~km~s$^{-1}$, the jet kinetic luminosity is $\sim 3\times
10^{36}$ erg s$^{-1}$, more than three orders of magnitude larger than
the non-thermal luminosity. Moreover, the energy density of the shocked 
thermal gas should be about two orders of magnitude
larger than that of the non-thermal particles. Therefore, to sustain
the non-thermal emission estimated in this work under reasonable
assumptions seems not to be a problem from the energetic point of view.

Summarizing, the fact that synchrotron radio emission has been
detected from IRAS~16547$-$4247 already indicates that there is
conversion of energy of some sort to non-thermal particles, and the
high densities present in the region, as well as the relatively high
kinetic luminosity of the jet, render likely that radiation will be
significantly generated by channels other than synchrotron radiation,
like relativistic Bremsstrahlung or even $pp$ interactions. After 
studying in a first-order approach
a reasonable scenario for the non-thermal emitting region, we conclude
that YSOs like IRAS~16547$-$4247 could produce detectable non-thermal 
emission via
synchrotron, relativistic Bremsstrahlung and $pp$ interactions from
radio up to very high-energies. This opens a new window for studies of
massive star formation in the near future.

\begin{acknowledgements}
A.T.A. and G.E.R. are supported by CONICET (PIP 5375) and the
Argentine agency ANPCyT through Grant PICT 03-13291 BID 1728/OC-AC.
V.B-R. thanks the Max-Planck-Institut f\"ur Kernphysik for its support
and kind hospitality.  V.B-R., and J.M.P  acknowledge support by DGI
of MEC under grant AYA2004-07171-C02-01, as well as partial support by
the European Regional Development Fund (ERDF/FEDER).
V.B-R. gratefully acknowledges support from the Alexander von Humboldt
Foundation.
\end{acknowledgements}
{}

\begin{thebibliography}{}

\bibitem{} Aharonian, F.A., Atoyan, A.M., 2000, A\&A, 362, 937 
\bibitem{} Bell, A.R., 1978, MNRAS, 182, 147
\bibitem{} Blumenthal, G.R., Gould, R.J., 1970, Rev. Mod. Phys., 42, 237
\bibitem{} Bonnell, I.A., Bate, M.R., Zinnecker, H., 1998, MNRAS, 298, 93
\bibitem{} Bosch-Ramon, V., Romero, G.E., Paredes, J.M., 2005, A\&A, 429, 267
\bibitem{} Brooks, K., Garay G.,  Mardones, D., Bronfman, L., 2003, ApJ, 594, L131 
\bibitem{} Garay, G., Brooks, K., Mardones, D., Norris, R.P., 2003, ApJ, 537, 739
\bibitem{} Ginzburg, V.L., Syrovatskii, S.I., 1964, The Origin of Cosmic Rays, Pergamon Press, New York
\bibitem{} Khangulyan, D., Hnatic, S., Aharonian F., Bogovalov S. 2007, MNRAS, in press
\bibitem{} Kelner, S.R., Aharonian, F.A., \& Vugayov, V.V., 2006, Phys. Rev. D, 74, 034018
\bibitem{} Mannheim, K., Schlickeiser, R., 1994, A\&A, 286, 983 
\bibitem{} Mart{\'\i}, J., Rodr{\'\i}guez, L.F., Reipurth, B., 1995, ApJ, 449, 184
\bibitem{} Pacholczyk, A.G., 1970, Radio Astrophysics, Freeman, San Francisco
\bibitem{} Pfrommer, C., En$\beta$lin, T.A., 2004, A\&A, 413, 17 
\bibitem{} Protheroe, R.J., 1999, in: Topics in Cosmic-Ray Astrophysics, 1999, p.247 [astro-ph/9812055]
\bibitem{} Rodriguez L.F., Garay G., Brooks, K., Mardones, D., 2005, ApJ 626, 953
\bibitem[]{}Romero, G.~E. 2001, in: The Nature of Unindentified Galactic
High-Energy Gamma-Ray Sources, ed. A. Carraminana, O. Reimer, \& D. Thompson, Kluwer
Academic Publishers, Dordrecht, 65
\bibitem{} Romero, G.E., Benaglia, P., Torres, D.F., 1999, A\&A, 348, 868
\bibitem{} Shu, F.H, Adams, F.C., Lizano, S. 1987, ARA\&A, 25, 23 

\end{thebibliography}
\end{document}